\documentclass[%preprint,
preprint,
 amsmath,amssymb,
 aps,
pra,
]{revtex4-1}

\usepackage{graphicx}% Include figure files
\usepackage{dcolumn}% Align table columns on decimal point
\usepackage{bm}% bold math

\begin{document}

\preprint{}

\title{Stored Ultracold Neutron Lifetime Experiments: \\Non-Inertial Frame Effects on the Neutron Velocity Spectrum }% Force line breaks with \\

\author{Steve K. Lamoreaux}
\affiliation{Physics  Department, Yale University\\ New Haven, Connecticut, U.S.A.}%Lines break %automatically or can be forced with \\
\email{Steve.Lamoreaux@yale.edu}%

\date{April 1, 2018}% It is always \today, today,

\begin{abstract}
Ultracold neutrons (UCN), stored in a cell with hard walls that is attached to, and therefore rotating with, the Earth, will experience
non-inertial frame effects, resulting in a broadening of the UCN spectrum. A heating or cooling of the spectrum is also possible.  This is because the stored UCN
are in a freely falling (inertial) frame between sudden wall collisions, and the acceleration of the cell relative to
this frame means that the cell walls will have a new quasi-random velocities between subsequent UCN wall collisions.  This results in a random
walk of UCN trajectories in momentum space.  Estimates of the effects on UCN with specified initial velocities are presented. Although the effects appear as small, they are worth considering for experiments of current interest, and will be important for possible future experiments with anticipated improved accuracy.
\end{abstract}
%\begin{description}
%\end{description}
%\end{abstract}
%
%\pacs{none}% PACS, the Physics and Astronomy
%                             % Classification Scheme.
%%\keywords{Suggested keywords}%Use showkeys class option if keyword
%                              %display desired
\maketitle

%\tableofcontents

\section{Introduction}

Although gravitation is by far the dominant force affecting the motion of ballistic objects, the effects of the Earth's rotation on ballistic trajectories have been appreciated since at least the mid-17th century. Earth rotation effects leads to an important correction the aiming of long-range artillery, and such corrections also apply to sharp-shooters' (snipers') aiming for distances larger than a thousand meters, depending on direction and latitude.

Such effects are also at play in stored Ultracold Neutron (UCN) lifetime experiments, and apparently have not been considered before, at least not in any great detail.  The purpose of this note is to estimate the magnitude of these effects. It will be shown that they are large enough to be of possible concern at the present level of accuracy for the current state of the art.

Consider the coordinate system in Fig. 1, which specifies the location and motion of a cubical storage cell as shown in Fig. 2, near the Earth's surface. The cell bottom lies in the hatched disk of Fig. 1, and the direction between the AB sides, taken as the $x$ axis, is parallel to that disk and intersects the $NS$ axis.
\begin{figure}[h]
\includegraphics[width=2in]{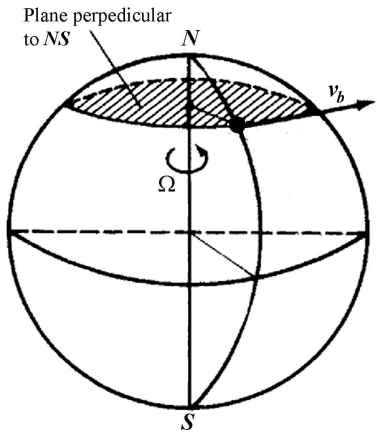}% Here is how to import EPS art
\caption{Coordinate system used to describe the storage vessel orientation and motion. Adapted from \cite{1}.}
\end{figure}

\begin{figure}[h]
\includegraphics[width=4in]{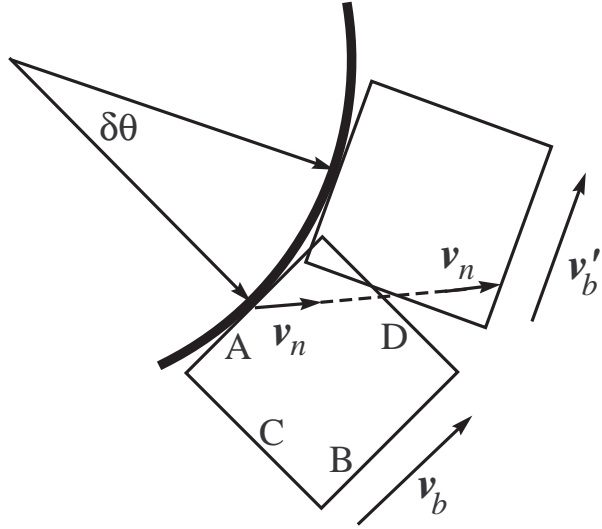}% Here is how to import EPS art
\caption{ A hypothetical cubical UCN storage cell which is located near the Earth's surface. The bottom lies in the hatched plane shown in Fig. 1, so to first order only the two sides A and B have a
change in velocity with rotation through a small angle $\delta\theta$. Note that the cell top and bottom are not perpendicular to the Earth's gravitational field $\vec g$, which is at an angle $\theta_L$ (the latitude) to those surfaces, with A and B at the complementary angle. However C and D are parallel to $\vec g$.}
\end{figure}

Let us imagine an enormous cell where no wall collisions occur for long time, e.g., the neutron lifetime
(neglect the Earth's gravitational field for the time being).  If an UCN is introduced in the cell's rest frame at time $t_0=0$, at a later time $t$ the A and B wall of the cell will be moving at a velocity along $x$ relative to the original reference frame ($v_{bx}=0$), to first order,
\begin{equation}
v_{bx}'=-v_b\delta\theta=-v\Omega t
\end{equation}
where $v_b$ is the azimuthal velocity,
\begin{equation}
v_b=\Omega r_e \cos\theta_L \approx 300\ {\rm m/s},
\end{equation}
$\Omega=2\pi/86400 {\rm s} = 7.27\times 10^{-5} {\rm rad\ s^{-1}}$ is the angular rotation rate of the Earth, $r_e$ is the Earth radius, and $\theta_L\approx 45^\circ$ is the approximate latitude where most storage cell experiments have been performed.
Then, in 900 s, the B wall will be moving with velocity, in the $x$ direction, of
\begin{equation}
v_w=-300{\rm\ m/s}\times 7.27\times 10^{-5} {\rm rad\ s^{-1}}\times 900\ {\rm s}=-19.6\ {\rm m/s}
\end{equation}
which is too fast to contain UCN, usually defined as having velocity magnitude relative and normal to a wall less than 7 m/s.

UCN stored in a more reasonable sized cell will undergo many wall collisions over a long storage period. Given that the change velocities of the A wall and of the B wall are opposite as far as UCN reflections are concerned, subsequent reflections tend to cancel the the UCN velocity changes.  This cancelation is not perfect due to a number of factors, including non-specular reflections which cause the time between wall collisions $\tau_c$ to randomly vary, fluctuations in $v_x$ between subsequent collisions with the A and B, due both to parabolic motion in the Earth's gravitation field and the non-specularity of wall reflections.  We can therefore model the evolution of the UCN spectrum as a random walk in velocity (momentum) space.

\section{Analysis}

When a UCN reflects from a wall moving with a velocity $v_w$ directed normal to the wall's surface, the component of the UCN velocity along the surface normal will change by
\begin{equation}
v_\perp'=v_\perp \pm 2v_w
\end{equation}
where the sign depends on the whether the wall in moving toward ($+$) or away from ($-$) the UCN. This of course assumes that $|v_w+v_\perp|< 7$ m/s otherwise the UCN would be lost.  It is well-know that wall motions parallel to the wall surface, in the absence of low energy resonances, do not affect the UCN velocity or wall reflection properties (for a perfect homogeneous plane surface) \cite{2}.

The change in the magnitude of the velocity is, for $v_w$ along $x$,
\begin{equation}
v_n'=\sqrt{(v_{nx}\pm 2v_w)^2+v_{ny}^2+v_{nz}^2}\approx v_n \sqrt{v_{nx}^2+v_{ny}^2+v_{nz}^2\pm 4v_{nx}v_w}\approx \left[v_n\pm 2 {v_{nx}\over v_n}v_w\right].
\end{equation}

For the case of a cell rotating with the Earth, because UCN reflections from the A or B walls occur at separated times, each reflection (collision) results in a change in UCN velocity (magnitude) due to the change in the velocities of the A and B walls during the time between collisions.  After each collision, we can transform into a new frame where the storage cell is at rest; in this rest frame, the UCN velocity will go through a quasi-random walk in velocity or momentum space.  This walk is not truly random because, for example, after a reflection from the A wall, a subsequent reflection from the B wall is more likely than another from the A wall (in the high velocity limit where gravity can be ignored). Nonetheless we can model this as a true random process, due to non-specularity of subsequent wall reflections and due to gravity which can cause $v_{nx}$ to vary between wall collision. In addition the time between wall collisions $\tau_c$  depends on the trajectory so is a quasi-random variable. This analysis works, of course, only works when the UCN energy is sufficiently high that reflections from B occur as frequently as reflections from A.

We can therefore define an approximate random walk length, noting that, from kinetic theory, $\langle v_{ny}\rangle =v_n/2$, as
\begin{equation}
\delta v_n= v_w= v_b\delta \theta= \Omega v_b \tau_c =r_e\cos\theta_L\Omega^2\tau_c
\end{equation}
where $\tau_c$ is the time between subsequent reflections from A and/or B.

If we start with a specific UCN velocity $v_n$, the velocity will spread due to these random processes.  The three-dimensional velocity (momentum) space diffusion coefficient can be written
\begin{equation}
D=6 {(\delta v_n)^2\over 3\tau_c}.
\end{equation}
The factor of 3 is included because only one third of the wall collisions result in a change of velocity.

The root-mean-square change in velocity, after a time $t$ is
\begin{equation}
\sqrt{\langle(\Delta v)^2\rangle}=
\sqrt{Dt\over 2}=r_e\cos\theta_L\Omega^2\sqrt{\tau_c\ t}=v_b\Omega\sqrt{\tau_c\ t}.
\end{equation}

Taking a typical $\tau_c=.03$ s,
\begin{equation}
\sqrt{\langle(\Delta v)^2\rangle}=3.8 \times 10^{-3} \sqrt{t}\ \ {\rm m/s}
\end{equation}
or, after $t=1000$ s, an initially single velocity UCN spectrum converts to a roughly Gaussian distribution with width $\pm 0.12$ m/s which is 4\% of a typical UCN velocity of 3 m/s, corresponding to an 8\% spread in energy. This is an effect that might be large enough to have an impact on experiments that have reported results better than $1\%$ accuracy.

Because the phase space volume is proportional to $v^2dv$, a spread in velocity results in a net larger average velocity because the higher velocities are more strongly weighted.  This is a well-known effect in a multi-dimensional random walk.

\section{Discussion and Conclusion}

An estimate is made of the expected random walk spread in velocity of UCN stored in a cell that is rotating with the Earth.  The system considered is highly idealized, and the effect for a real experiment will depend on many parameters, including the orientation of the storage cell relative to both the Earth's rotation axis and the gravitational field.

Because the effect scales at $\tau_c$ (in the high velocity limit considered so far), we might expect low velocity UCN to be affected more.  However, in the presence of gravity, $\tau_c$ becomes small as the UCN energy (velocity) goes to zero, because the UCN bounce as a ball on the lowest surfaces of the cell, with
\begin{equation}
\tau_c\approx 2 {v_n\over  g}
\end{equation}
and is largely independent of the cell dimensions.
For example, in Fig. 2, it is possible to have UCN with sufficiently low energy so that the B surface is never reached.  In this case, collisions with surface A always results in a decrease in UCN velocity that is never compensated by velocity-increasing collision with surface B; hence there exists a cooling process.  In addition, the net change in average velocity could scale as $t$ instead of $\sqrt{t}$, with the spread evolving as $\sqrt{t}$ but at a lower rate than the previous case.

Given the dependence on cell shape and orientation together with UCN velocity, general predictions on the effects on, and corrections to, the neutron lifetime as obtained from UCN storage experiments are not really possible.  The purpose of this note is to point out that corrections in the velocity spectrum of UCN stored in a cell can be modified significantly (at the few percent level) by the Earth's rotation, and should be included in a complete analysis of any experimental system.

The effects for soft-wall storage experiments, such a parabolic magnetic traps, are expected to be small because the Coriolis force can be combined with the trapping forces in a time-continuous manner.


\begin{thebibliography}{99}
\bibitem{1} The Great Soviet Encyclopedia, 3rd Edition. S.v. ``Coriolis acceleration." Retrieved April 1 2018 from https://encyclopedia2.thefreedictionary.com/Coriolis$+$acceleration

\bibitem{2} M.A. Horne, A. Zeilinger, A.G. Klein, and G.I. Opat, Phys. Rev. A {\bf A28}, 1 (1983).

\end{thebibliography}
\end{document}